# Massive topological edge channels in three-dimensional topological materials induced by extreme surface anisotropy

Fengfeng Zhu,[1, 2, 3, *] Chenqiang Hua,[4, 5, *] Xiao Wang,[3] Lin Miao,[6] Yixi Su,[3] Makoto Hashimoto,[7] Donghui Lu,[7] Zhi-Xun Shen,[7, 8] Jin-Feng Jia,[1, 9, 10, 11, 12] Yunhao Lu,[4, †] Dandan Guan,[1, 9, 10, 11, ‡] and Dong Qian[1, 9, 12 §]

[1]*Key Laboratory of Artificial Structures and Quantum Control (Ministry of Education), School of Physics and Astronomy, Shanghai Jiao Tong University, Shanghai 200240, China*

[2]*State Key Laboratory of Information Functional Materials, 2020 X-Lab, Shanghai Institute of Microsystem and Information Technology, Chinese Academy of Sciences, Shanghai, 200050, P. R. China*

[3]*Jülich Centre for Neutron Science (JCNS) at Heinz Maier-Leibnitz Zentrum (MLZ), Forschungszentrum Jülich, Lichtenbergstrasse 1, D-85747 Garching, Germany*

[4]*School of Physics, Zhejiang University, Hangzhou 310027, China*

[5]*Zhongfa Aviation Institute of Beihang University, Hangzhou 310023, China*

[6]*School of Physics, Southeast University, 211189 Nanjing, China*

[7]*Stanford Institute for Materials and Energy Sciences, SLAC National Accelerator Laboratory, Menlo Park, CA 94025, USA*

[8]*Geballe Laboratory for Advanced Materials, Department of Physics and Applied Physics, Stanford University, Stanford, CA 94305, USA*

[9]*Tsung-Dao Lee Institute, Shanghai Jiao Tong University, Shanghai 200240, China*

[10]*Hefei National Laboratory, Hefei 230088, China*

[11]*Shanghai Research Center for Quantum Sciences, 99 Xiupu Road, Shanghai 201315, China*

---

[*] These authors contributed equally to this work.
[†] luyh@zju.edu.cn
[‡] ddguan@sjtu.edu.cn
[§] dqian@sjtu.edu.cn

*[12]Collaborative Innovation Center of Advanced Microstructures, Nanjing University, Nanjing 210093, China*

## Abstract

A two-dimensional quantum spin Hall insulator exhibits one-dimensional gapless spin-filtered edge channels allowing for dissipationless transport of charge and spin. However, the sophisticated fabrication requirement of two-dimensional materials and the low capacity of one-dimensional channels hinder the broadening applications. We introduce a method to manipulate a three-dimensional topological material to host a large number of one-dimensional topological edge channels utilizing surface anisotropy. Taking $ZrTe_5$ as a model system, we realize a highly anisotropic surface due to the synergistic effect of the lattice geometry and Coulomb interaction, and achieve massive one-dimensional topological edge channels -- confirmed by electronic characterization using angle-resolved photoemission spectroscopy, in combination with first-principles calculations. Our work provides a new avenue to engineer the topological properties of three-dimensional materials through nanoscale tunning of surface morphology and opens up a promising prospect for the development of low-power-consumption electronic nano devices based on one-dimensional topological edge channels.

The quantum spin Hall insulator (QSHI) is a two-dimensional (2D) quantum system that hosts topologically protected one-dimensional (1D) edge states, making it a highly promising candidate for developing dissipationless devices[1–11]. A naive approach to enhance the quantity of 1D channels is to stack multiple QSHIs together, forming a bulk material[5,6]. However, in real crystalline materials, the interlayer coupling in stacked QSHIs leads to the emergence of three-dimensional (3D) weak or strong topological insulators (WTIs or STIs)[9,10,12]. Consequently, the 1D edge states transition into 2D surface states, resulting in an increase in the number of conducting channels and an elevated susceptibility to scattering. Thus, the non-dissipative characteristics of the system are significantly compromised. To restore or enhance the dissipationless properties, various proposals have been proposed, including increasing the interlayer distance or introducing insulating layers to preserve the 1D nature of the edge channels[4,5,9,10]. However, none of these proposals have been experimentally realized thus far.

We propose a novel strategy to manipulate a 3D topological material and induce a multitude of 1D topological edge channels through extreme surface anisotropy. We illustrate this concept in Fig. 1a, using a 3D topological insulator (TI) as a model system. According to the Bernevig-Hughes-Zhang and Kane-Mele model[1-3], the 2D Dirac state of 3D TI is described by $E^2 = v_x^2 k_x^2 + (1/A)v_y^2 k_y^2$, where $v_x(v_y)$ is the Fermi velocity along the x(y) direction and $A$ is the uniaxial anisotropy parameter. In the presence of an isotropic surface (Fig. 1a-left), i.e., $v_x = v_y$ and $A = 1$, a symmetric 2D Dirac cone emerges. However, by introducing an anisotropic surface characterized by atomically irregular 1D carved structures along the x-direction, as depicted in Fig. 1a-right, we can induce a uniaxial anisotropy. As the degree of $A$ increases, the initially isotropic Dirac cone gradually transforms into an anisotropic state, eventually giving rise to a 1D surface state with an abundance of conducting channels when $A$ becomes infinite (Fig.

1a-right). Fig. 1b shows a sketch of a 3D topological material with numerous 1D spin-momentum locked edge channels in the real space, as proposed.

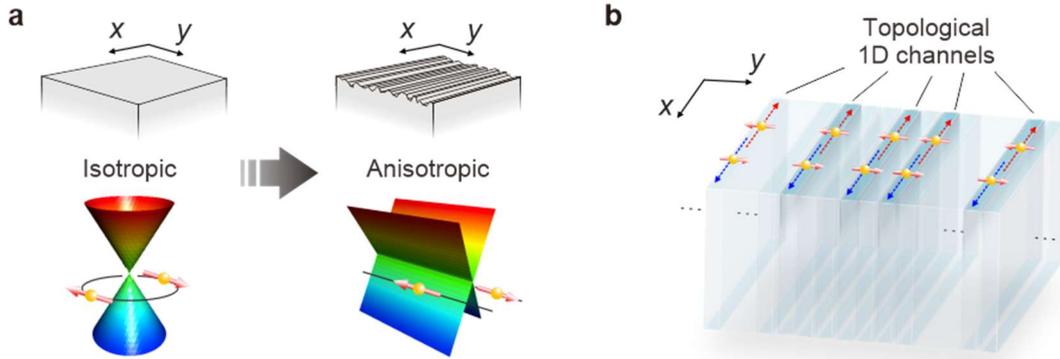

**Figure 1| Modulation effects of surface morphology on the anisotropy of a 2D Dirac cone state. a,** Surface morphology modulation and the corresponding change of Dirac cone states. When the smooth surface is substituted by abundant 1D micro-structures along the x-direction, the Dirac state of 2D surface gradually exhibits more 1D characteristics and finally becomes the 1D surface Dirac state with massive 1D conducting channels. **b**, A schematic plot of the massive 1D topological edge channels on the surface of a 3D TI material.

In practical experiments, utilizing a 3D STI may not be ideal due to the necessity of modifying all surfaces, as each surface of a 3D STI possesses a 2D Dirac cone. In contrast, a 3D WTI provides a more favorable option, where the topological surface state manifests only on specific surfaces[12–18]. The crucial challenge lies in achieving a highly anisotropic surface configuration. In this study, for the first time, we are able to generate a profusion of topologically protected, highly quasi-one-dimensional (Q1D) channels within a 3D WTI through the extremely anisotropic surfaces of $ZrTe_5$.

Single crystalline $ZrTe_5$ is a van der Waals (vdW) layered material with each layer stacked along the *b*-axis (Fig. 2a). One $ZrTe_5$ layer consists of $ZrTe_3$ and zigzag-Te chains along the *a*-axis. Single-layer $ZrTe_5$ is a 2D TI[13,19] and bulk $ZrTe_5$ can be a 3D WTI or STI depending on the lattice parameter according to the calculations[13,20,21]. The electronic properties of $ZrTe_5$ crystals have been extensively studied by ARPES[18,22–25], STM[19,26,27], magneto-transport measurements[28–41] and so on, which have revealed that the $ZrTe_5$ crystal is a 3D WTI. Recently, signatures of a Q1D edge state were observed in ZrTe$_5$, which further supports that ZrTe$_5$ is a WTI[18,19,26,27]. However, the theoretical model cannot well describe the band dispersion of the observed Q1D states[13,18].

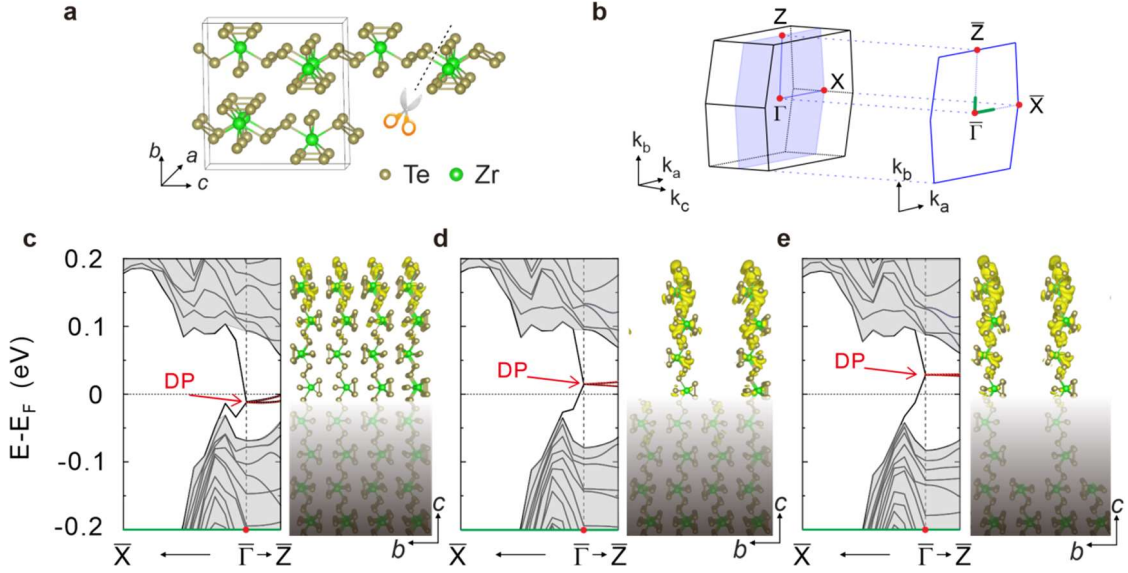

**Figure 2| Calculated band structures of ZrTe$_5$ with various surface morphology.** **a**, Lattice structure of bulk ZrTe$_5$. **b**, The bulk BZ and the projected surface BZ of the *a-b* plane. **c-e**, Representative surface-modification of the *a-b* plane and the corresponding band-structure. The corresponding momentum range is indicated by the green lines in **b**. The $\bar{\Gamma} - \bar{Z}/\bar{\Gamma} - \bar{X}$ directions respectively correspond to the *b*/*a* directions in **a**. From **c** to **e**, the surface Dirac state becomes more flat along the $\bar{\Gamma} - \bar{Z}$ direction, indicating the evolution from a 2D surface Dirac state to a 1D surface Dirac state.

Experimentally, ZrTe$_5$ can be cleaved along two planes, e.g., the *a-c* and *a-b* planes (see Fig. 2a). The cleavage of the *a-c* plane occurs within the vdW gap, which gives a flat surface without the topological surface state[18,23,24]. The cleavage of the *a-b* plane where the topological surface state locates is different. The cleavage of the *a-b* plane can break either Te-Te bonds or Zr-Te bonds. Currently, its surface condition is barely known. Using DFT-based first-principles calculations, we find that Te-Te bonds are stronger than Zr-Te bonds (Extended Data Figs. 1 and 2), and thus Zr-Te bonds break more easily during cleavage. With the breaking of bonds, one must consider the additional electrostatic Coulomb interaction at the edges. To minimize the Coulomb energy, the *a-b* plane cleavage will not favor a smooth surface, i.e., forming a configuration of groove-like edges between the adjacent ZrTe₅ layers, which will effectively reduce the interlayer coupling at the edge. Fig. 2b shows the bulk and the *a-b* surface Brillouin zone (BZ). We demonstrate the effects of groove-like edges on the surface state in Figs. 2c-e. The atomic models in Figs. 2c-e show the side view along the *a*-axis. Bands shown in shadow represent the bulk band projections. Within the bulk energy gap, a topological surface state appears. The Dirac point (DP) of the surface band is indicated by the red arrow. For a flat surface (Fig. 2c), the surface state is an anisotropic 2D Dirac state. The Fermi velocity is larger along the $\bar{\Gamma} - \bar{X}$ direction than along the $\bar{\Gamma} - \bar{Z}$ direction. When a groove-like nano structure forms, the surface state becomes more anisotropic. The surface Dirac state becomes nearly nondispersive along the $\bar{\Gamma} - \bar{Z}$ direction, forming a 1D Dirac surface state when the anisotropy is large enough (see in Fig. 2e). The 1D Dirac surface state on a groove-like surface is equivalent to a simple combination of many decoupled QSHI's edge states, as demonstrated in Fig. 1b.

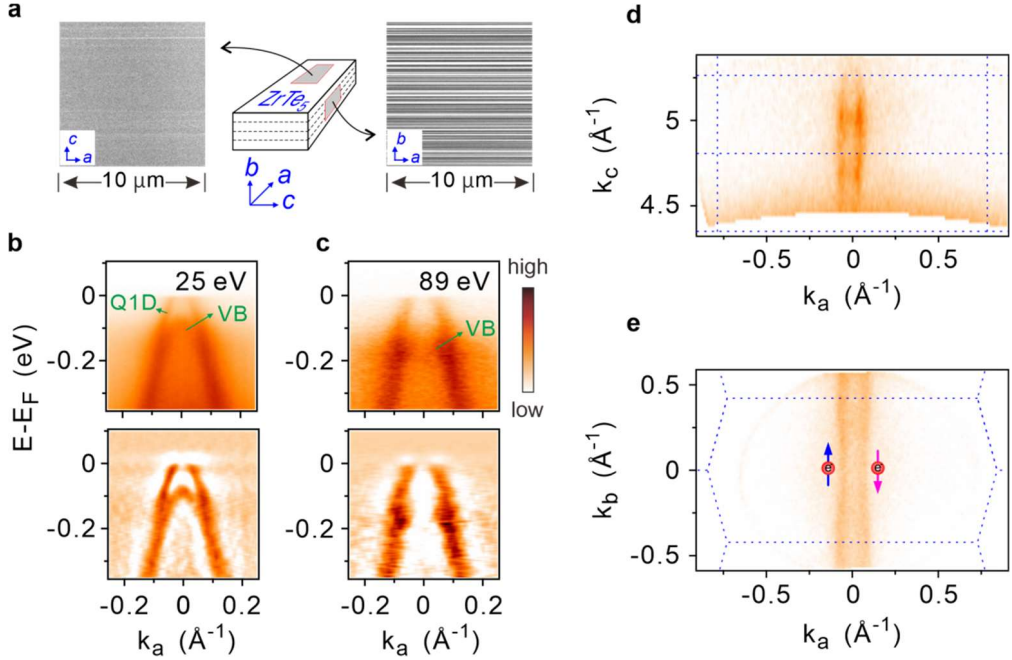

**Figure 3| ARPES spectra of the Q1D surface state of ZrTe₅ on the extremely anisotropic surface. a,** SEM images of the fresh cleaved surface for the *a-c* (left) and *a-b* plane (right). Enormous 1D nano structure exists on the *a-b* plane. **b** and **c,** ARPES

band dispersions and the corresponding second derivative plots along the $k_a$ direction measured on the *a-b* surface at 10 K with two incident photon energies ($hv$ = 25 eV and 89 eV). The Q1D surface state and bulk VB are indicated by arrows. The Fermi surface of the Q1D surface state was measured **d,** in the $k_a$-$k_c$ plane by photon energy dependence and **e,** in the $k_a$-$k_b$ plane with incident photons of $hv$ = 23 eV. The blue dashed lines in (**d**) and (**e**) indicate the surface BZ boundaries. Two arrows in (**e**) indicate the spin orientations of the electrons on the Fermi surface.

To verify our scenario, the *a-c* and *a-b* surfaces were imaged with scanning electron microscopy (SEM). In Fig. 3a-left, as expected, the cleaved *a-c* surface is smooth. Meanwhile, the *a-b* surface (Fig. 3a-right) shows completely different morphology. Within our SEM resolution, a high density of parallel 1D nano structures along the *a*-axis were observed. The synergistic effect of the chain-like structure along the *a*-axis in the lattice and the Coulomb interaction on the edge forms the 1D nano structure. More details about the surface morphology can be found in Extended data Fig. 3. The 1D nano structure on the surface very likely provides a strong surface anisotropy to manipulate the surface state.

We conducted high resolution ARPES measurements on the *a-b* surface to check the consequence of the 1D nano structure. Fig. 3b shows the ARPES spectra of ZrTe$_5$ near E$_F$ along the $k_a$ direction using an incident photon energy ($hv$) of 25 eV. A linearly dispersive hole-like band crossing the E$_F$ was clearly resolved, and another hole-like band below the E$_F$ was also detected. This lower hole-like band is a bulk valence band (VB) that moves away from the E$_F$ when we change incident photon energy (corresponding to the change in $k_c$) to $hv$ = 89 eV (Fig. 3c). The position of the upper hole-like band does not move when we change $k_c$. Figs. 3d and 3e show the Fermi surface of the upper hole-like band in two orthogonal planes in momentum space. It has no dispersion along the $k_c$ direction (Fig. 3d) and is very slightly dispersive along the $k_b$ direction (Fig. 3e), indicating that this linearly dispersive hole-like band comes from a Q1D state. Spin-resolved ARPES experiments were performed to further elucidate the spin configuration of this Q1D state in momentum space. These experiments confirm the behavior of spin-momentum locking as expected in spin-orbit coupled systems. The spin polarization is found to be parallel to the $k_b$ direction, as indicated in Fig. 3e. Details of the spin configuration can be found in Extended Data Fig. 4. Overall, the band dispersion and spin configuration of the detected Q1D states are consistent with the previous experimental results[18,25]. However, the real nature of these Q1D states cannot be revealed unless the groove-like 1D surface structure is considered. In fact, the Q1D states on the *a-b* surface are not the intrinsic surface state of a 3D WTI, but a unique surface state consisting of many 1D edge state of QSHI. We will discuss this in Fig. 4.

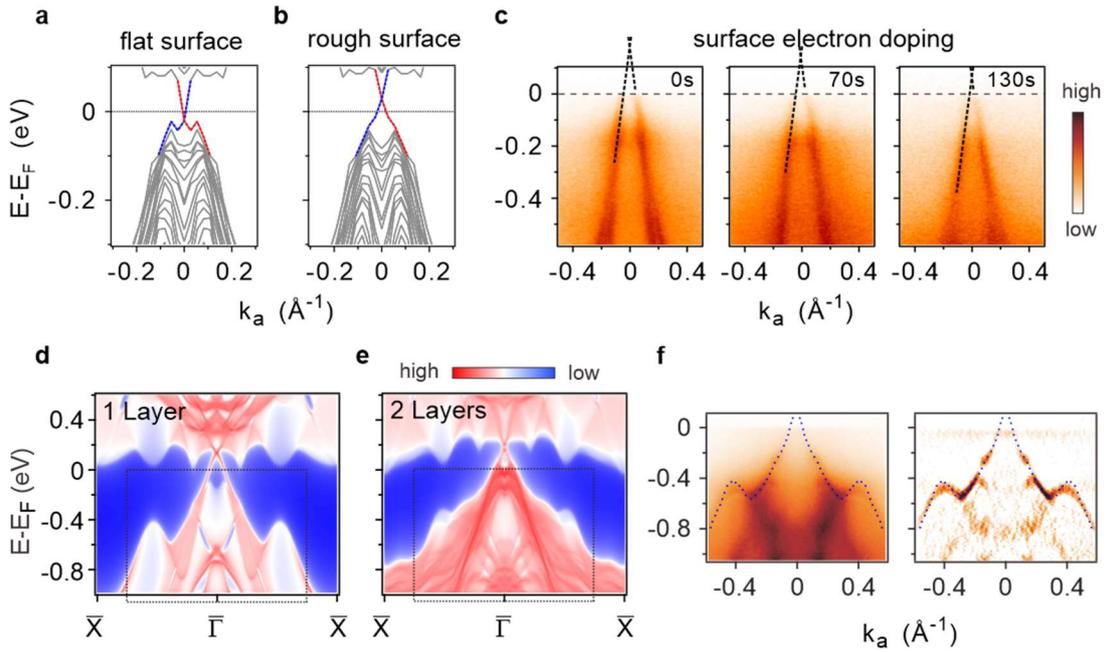

**Figure 4| Comparison between the experimental band dispersion and the DFT calculations of the Q1D state considering the 1D nano structure in ZrTe$_5$. a** and **b**, The calculated band dispersions on a flat and rough *a-b* surface, respectively, of bulk ZrTe$_5$ with Te termination by breaking Zr-Te bonds. **c**, ARPES spectra as a function of surface potassium deposition time. The black dashed lines are guidelines for visualization. **d** and **e**, The calculated band dispersions along the $\bar{X} - \bar{\Gamma} - \bar{X}$ direction of semi-finite single-layer and bi-layer ZrTe$_5$, respectively, with Te termination via the breaking of Zr-Te bonds. The 1D edge state exists within the 2D bulk band gap. **f**, ARPES spectra (left) and the corresponding second derivative plot (right) along the $k_a$ direction, measured within the energy-momentum range of the black dotted rectangle in (**d**). The blue dotted lines in (**f**) represent the band dispersion extracted from the calculated 1D edge state in (**d**).

We present the calculated low energy bands along the $k_a$ direction crossing the $\bar{\Gamma}$ point on both a flat and rough (with 1D nano structure) *a-b* surface with the breaking of Zr-Te bonds in Figs. 4a and 4b, respectively. For a rough surface, periodic 1D surface structure such as that in Fig. 2e is used. In all calculations, the surface is Te-terminated. The solid red/blue curves represent the surface state. Different colors indicate the opposite spin orientation. The dispersions of the surface bands from the two kinds of surfaces are very different. For the flat surface (Fig. 4a), the DP is below the $E_F$ and the surface band has a "W" shape dispersion below the DP. In contrast, the surface band is much simpler for the rough *a-b* surface. It shows an "X" shaped dispersion (Fig. 4b) and the DP is above the $E_F$. The experimentally detected surface band (Fig. 3b) clearly disagrees with the calculated surface band on the flat surface, but is similar on the rough surface. We also calculated band structures with the breaking of Te-Te bonds (Extended Data Fig. 5), which shows a completely different behavior from the experimental results. Additionally, we carried out *in-situ* surface potassium deposition (surface electron doping) to check the band dispersion above the $E_F$. As shown in Fig. 4c, the bands gradually shift downwards with potassium doping. After 130 seconds of potassium deposition, although the DP has not been reached, the surface band maintains a simple linear dispersion, which further supports that the detected Q1D states are very likely from a rough surface with 1D nano structure.

Note that although periodic 1D nano structure is used in the calculation, some randomness is expected on the real cleaving surface. Nevertheless, with many groove-like 1D nano structures along the *a*-axis, the interaction between adjacent $ZrTe_5$ layers could be significantly weakened near the edges. Therefore, we believe that the band dispersion of the surface state on the *a-b* surface with 1D nano structure could be analogous to the dispersion of the edge state of a single $ZrTe_5$ layer. To check this, we calculated the energy band of a semi-finite single-layer of $ZrTe_5$ with the breaking of Zr-Te bonds on the edge, as shown in Fig. 4d. For comparison, the band structure of a semi-infinite double-layer $ZrTe_5$ is also presented in Fig. 4e, which is very different from the single-layer $ZrTe_5$. Te-termination is used in the calculation. Band structures of single-layer $ZrTe_5$ with different edge terminations can be found in Extended Data

Fig. 6. For a single-layer ZrTe$_5$, there is a simple linearly dispersive band within the bulk gap near the $\bar{\Gamma}$ point, and the DP is located above the E$_F$. The calculated band dispersion of the edge state (blue dotted lines from Fig. 4d) is overlayed on the ARPES spectra and the corresponding secondary derivative intensity plot in Fig. 4f. Astonishingly, the agreement between the calculations and the experiments is nearly perfect despite we measuring on a 3D material, which indicates that the extremely anisotropic surface morphology does tune the 2D topological surface state into a QSH-like topological edge state with a very high density of 1D channels in ZrTe$_5$.

As a 3D WTI, ZrTe$_5$ only has a topological surface state in the *a-b* and *b-c* planes. When we apply currents along the *a*-axis (electrons move in the *a-b* surface), the massive topological 1D edge channels generated in the *a-b* surface have the capacity to carry high charge or spin currents. In principle, this method of utilizing the synergistic effect of the lattice structure and the Coulomb interaction to induce extreme surface anisotropy can be used not only in material analogues of ZrTe$_5$, like HfTe$_5$, but also in other layered topological materials with a chain-like structure. Our strategy to realize massive 1D channels in a 3D topological material moves a tremendous step forward towards the future application of the topological 1D edge state.

# METHODS

## First-principle calculations

The Vienna ab initio simulation package (VASP)[42] was used to perform the first-principles calculations and the projected augmented wave (PAW) method was employed for the description of the ion-electron interactions[43]. The exchange-correlation functional was modeled by the Perdew-Burke-Ernzerhof (GGA-PBE) formulism[44]. An energy cutoff of 500 eV was adopted for the plane-wave basis. The convergence thresholds of energy and force were set to $1\times10^{-6}$ eV and 0.01 eV/Å, respectively. Spin-orbit coupling was included to catch the band-topology[45]. To obtain the topological states of the semi-infinite structure, Wannier functions based on the p-orbital of Te were constructed[46,47]. To simulate the 1D/2D slab models, a ~40/20 Å vacuum region was adopted to avoid the unphysical interactions between neighboring slabs. The Brillouin zones of the $ZrTe_5$ primitive cell, 2D, and 1D ribbon systems were sampled by the $11 \times 11 \times 7$, $11 \times 7 \times 1$, and $11 \times 1 \times 1$ Γ-centered k-mesh, respectively. The crystal orbital Hamilton population (COHP) with a plane-wave basis was calculated to describe the bonding information between atoms[48].

## Sample growth

High-quality single crystals of $ZrTe_5$ were grown by the chemical vapour transport method in a three-zone furnace. The raw materials of Zirconium slug (99.95%) and Tellurium powder (99.99%) were sealed together with the high purity transport agent Iodine (>99.998%) in an evacuated silica ampoule 12 cm in length and 2 cm in diameter. During the first stage, the ampoule was placed into a horizontal furnace that was heated to 450 °C for the source zone and 500 °C for the growth zone and kept there for 10 days. This pre-grown process is to disintegrate the Zr slug to form $ZrTe_5$ polycrystals in the source zone and also clean the inner surface of the silica ampoule in the vicinity of the growth zone. Afterward, the temperatures of the source zone and growth zone were swapped to start the crystal growing process and sustained for another 10 days. Shiny and stick-like crystals with a typical dimension of about 6 mm × 1.5 mm × 1 mm were obtained.

**ARPES measurements**

Synchrotron-based ARPES measurements were performed at the Stanford synchrotron radiation light source (SSRL) beamline 5-2, the advanced light source (ALS) beamlines 4.0.3 and 10.0.1, and the national synchrotron radiation laboratory (NSRL) beamline 13U. Photoelectrons were detected with a Scienta Omicron R4000 analyzer and a DA30L analyzer. The proper incident photon energy was selected within 20 eV to 120 eV, according to the spectra intensity at different ARPES beamline stations. The base temperature of the samples during measurements is set to 20 K at NSRL and ALS, and 10 K at SSRL. The spin-resolved ARPES experiments were performed at NSRL beamline 13U and the spin-polarizations of photoelectrons were detected by two very-low-energy electron diffraction (VLEED) spin detectors which can detect two spin components, where one component is in-plane and parallel to the ARPES slit and the other is out-of-plane.

**Bond breaking and surface termination**

To determine the surface of $ZrTe_5$ when cleaving along the *a-b* plane, the electron localization function (ELF) is calculated to capture the charge distribution. Shown in 2D ELF (Extended Data Fig. 1a), red and blue colors indicate the electron accumulation and depletion, respectively. Clearly, more electrons are distributed between Te-Te than between Zr-Te, indicating the considerably stronger bonding within the Te-Te bonds. The COHP is also exploited to explore the bonding strength (see Extended Data Fig. 1b). Consistent with Extended Data Fig. 1a, the almost negative -pCOHP results of the Zr-Te bonds denotes the anti-bonding feature, manifesting the crumblier Zr-Te bonds relative to the zigzag Te chain. This can be manifested by the integrated -pCOHP value, i.e., Zr-Te is -4.4 and Te-Te is 0.3. Negative/positive results directly indicate the anti-bonding/bonding features of Zr-Te/Te-Te bonds, respectively, in agreement with the ELF calculation. Therefore, the Zr-Te bonds will break easily when exfoliating the *a-b* plane, leaving the remaining Te-Te dimers at the edge.

As depicted in Extended Data Fig. 2a, there are topological edge states localized on the outermost Te/Zr atoms, resulting in a possible Coulomb interaction between them.

Accordingly, we enumerate the typical surface configurations in Extended Data Fig. 2b after the bonds are broken, with three rough (I, II, and III) and two flat structures. Then, we calculate the relative energy in Extended Data Fig. 2c. The flat structures are higher in energy than the rough surfaces. With an increase in roughness, energy becomes lower by minimizing the repulsion, which again demonstrates that a rough surface is favorable during the exfoliation. It is further found that the II structure is the lowest-energy configuration with a fixed roughness, e.g., I and III are 0.35 and 0.15 meV/Å$^2$ higher than the II surface (see 2 u.c. in Extended Data Fig. 2c), respectively.

**Surface morphology of the cleaved *a-b* and *a-c* surfaces measured by SEM**

The surfaces of the cleaved a-b and a-c planes were checked by SEM under different magnification. Within the resolution of our SEM (spot size is ~2.4 nm, image pixel size is ~2.7 nm), two surfaces exhibit completely different surface morphology. In Extended Data Figs. 3a and 3b, the *a-b* surface shows a very high density of parallel 1D structures along the *a*- axis. However, on the *a-c* surface, except for some big steps at the edges, the whole surface is quite smooth, which conforms to the characteristics of the van der Waals layer, as shown in Extended Data Figs. 3c and 3d.

**Spin configuration of the Q1D band**

Our instrument has two spin channels: one being the in-plane channel (parallel to the analyzer slit) and the other being the out-of-plane channel, namely channel Y and channel Z, as shown in Extended Data Fig. 4a. To determine the spin configuration, we checked two ARPES spectra (Extended Data Fig. 4b) along the cut-1 and cut-2 directions, shown in Extended Data Fig. 4c. The orange solid lines in Extended Data Fig. 4c sketch the Fermi surface of the Q1D state. The spin asymmetry of the energy dispersion curves (EDCs) was measured at the Fermi vectors $k_1$ and $k_2$ (see Figs. 4a and 4b). The spin asymmetry (*P*) is defined by $P = (I_\uparrow - I_\downarrow)/(I_\uparrow + I_\downarrow)$, in which $I_\uparrow$ and $I_\downarrow$ represent the measured intensity of opposite spins. Extended Data Fig. 4c presents the spin asymmetry of the EDCs along the *a*-, *b*-, and *c*-axes. No sign of spin dichroism was observed along the *a*- or *c*-axis. Instead, spin dichroism was observed along the *b*- axis. We further measured the spin asymmetry of the momentum dispersion curve (MDC) at the binding energy of 48 meV, labeled by the green dotted line in Extended Data Fig. 4b. Consistent with the spin dichroism of the EDCs along the *b*-axis, the MDC exhibits a spin sign change along the *b*-axis, as shown in Extended Data Fig. 4e. The observed spin dichroism both in the EDC and MDC indicates that the spin of the Q1D state is along the *b*-axis and is spin-momentum locked, confirming its topological nontrivial nature. Our determined spin configuration is consistent with previous spin-resolved experiments.

## ACKNOWLEDGMENTS

We acknowledge Susanne Mayr, Xuxu Bai for the assistance with the crystals' alignment by X-Ray Diffraction, Guangzhou Ye for the assistance with ARPES experiments at SSRL, Shengtao Cui, Yuliang Li for the support during measurement of spin-ARPES experiments, Jie Ma and Guohua Wang for the help of heat capacity measurements. This work is sponsored by the National Key Research and Development Program of China, Shanghai Pujiang Program, the Strategic Priority Research Program of Chinese Academy of Sciences, the Science and Technology Commission of Shanghai Municipality, the Innovation program for Quantum Science and Technology, the Zhejiang Provincial Natural Science Foundation of China and the National Natural Science foundation of China. The Stanford work was supported by the US DOE Office of Basic Energy Sciences, Division of Materials Science and Engineering.

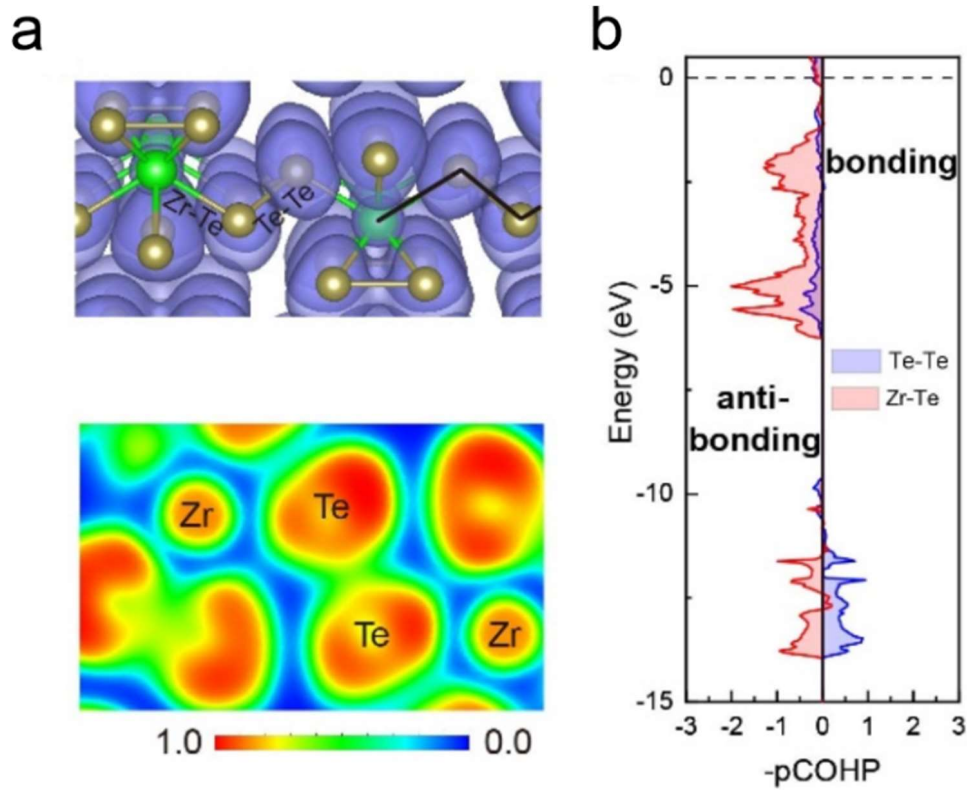

**Extended Data Fig.1**. **DFT determined bonding and surface when cutting the *a-b* plane of ZrTe₅**. **a**, The electron localization function (ELF) of ZrTe₅ structure. The black line denotes the slice plane of ELF in the bottom panel. Here 1 and 0 in the colorbar denote the electron accumulation and depletion, respectively. **b**, The calculated -pCOHP plots of different interactions when cutting the surface of ZrTe₅ (breaking Zr-Te and Te-Te bonds respectively). The positive/negative value denotes the bonding/anti-bonding, respectively. For example, the integrated -pCOHP value of Zr-Te bonds is much smaller than that of Te-Te bonds (ZrTe: -4.4; Te-Te: 0.3), indicating the considerably stronger bonding within the Te-Te bonds.

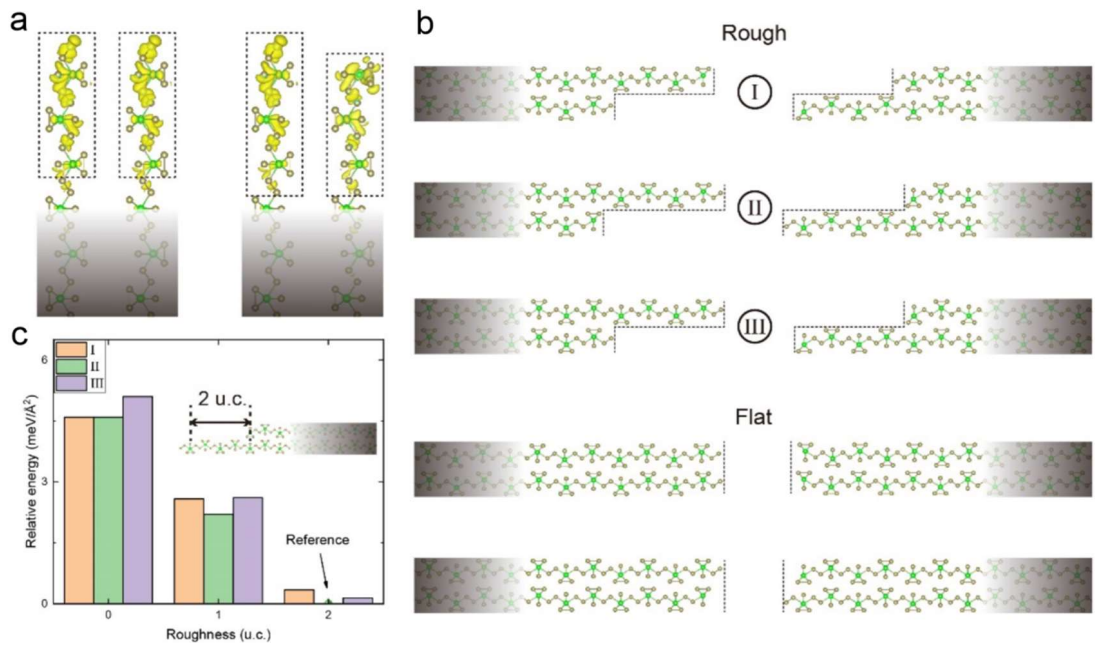

**Extended Data Fig.2. Surface configurations. a**, The schematic plots of the possible Coulomb interaction due to the Te/Zr terminations on the flat surface. The black dash line denotes the topological states at the near edge, partially contributing to the Coulomb interaction. **b**, The schematic plots of the typical configurations after breaking of Zr-Te bonds, i.e., rough surfaces (I to III) and flat ones. **c**, Energy comparison between different surface configurations. One u.c. represents that the displacement along the *c*-direction is one unit-cell length of 13.57 Å, causing the roughness. Flat surface is indicated by 0 u.c. See surfaces I to III in **b**. Inset is the structure of the reference point (0 meV/Å$^2$), i.e., II structure in **b**.

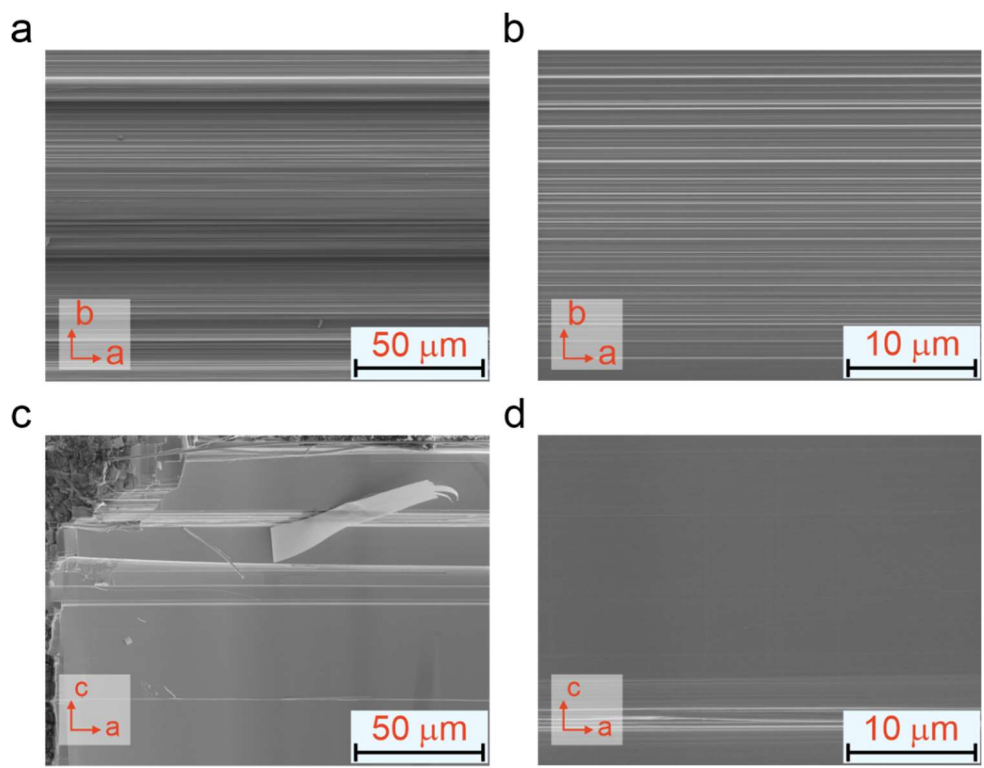

**Extended Data Fig.3. SEM images of the cleaved *a-b* and *a-c* surfaces under 5 keV electrons**. **a** and **b**, SEM images of the *a-b* surface with the magnification of 2000 and 20000. Groove-like 1D micro structures along the *a*-axis can be recognized. **c** and **d**, SEM images of the *a-c* surface with the magnification of 2000 and 20000. The cleaved surface is quite flat and smooth except for several big steps.

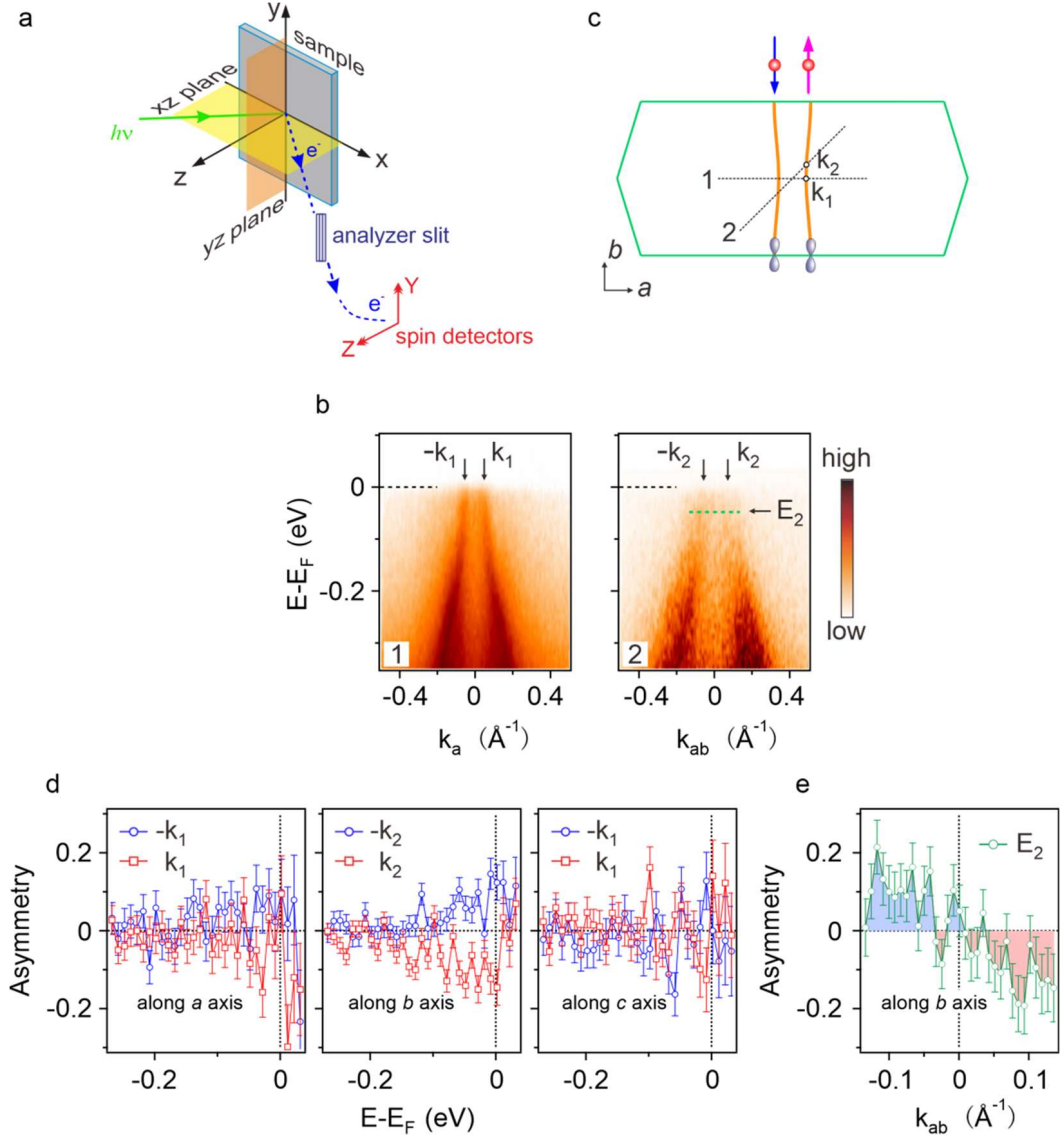

**Extended Data Fig.4. Spin configuration of the Q1D states. a**, A sketch for the Spin-ARPES experiment configuration of BL13U at NSRF. **b**, The ARPES spectra of the Q1D states measured along the line "1" and line "2" in **c** across the surface BZ center $\bar{\Gamma}$ point. **c**, A schematic plot of the Q1D states dispersion on the Fermi surface of the *a-b* plane. The orange and green solid lines represent the Q1D states and surface BZ boundaries respectively. Purple and blue arrows indicate the alternative spin of the electrons in the Q1D state. **d**, The reduced spin asymmetry of the energy dispersive curves (EDCs) at the momentum $k_1$ and $k_2$ in **b**1. **e**, The reduced spin asymmetry of the momentum dispersive curve (MDC) at energy $E_2$= 48 meV in **b**2 along the *b*-axis. The red and blue shadow represents the opposite spin polarization.

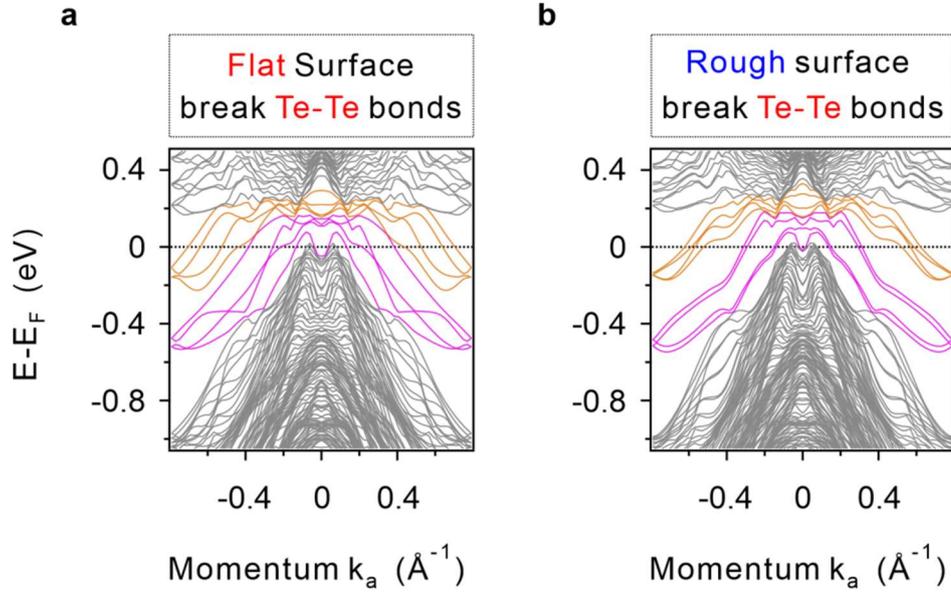

**Extended Data Fig.5. Calculated band dispersion on the *a-b* surface of ZrTe$_5$ assuming the breaking of Te-Te bonds. a,** Flat and **b**, Rough *a-b* surface of ZrTe$_5$ with Te-Te bonds breaking. The in-gap trivial and non-trivial surface states are indicated by the orange and purple lines respectively.

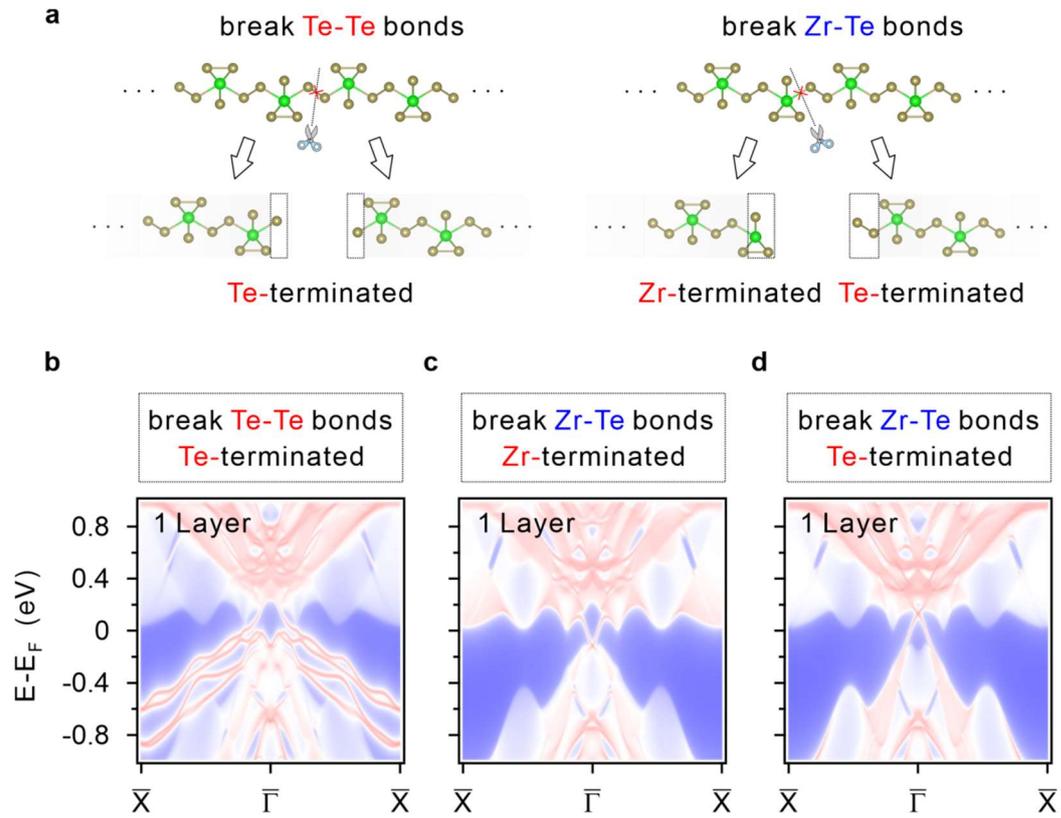

**Extended Data Fig.6. Calculated band dispersion of a semi-finite single-layer ZrTe$_5$ with different terminations. a**, A schematic plot for the various terminations at the edge of a single-layer ZrTe$_5$ after breaking different bonds. **b-d**, The calculated edge states of a semi-finite single-layer ZrTe$_5$ with various terminations and bonds breaking.